\documentclass[singlespacing]{elsart}
\usepackage{amssymb,amsmath}
\usepackage{graphicx}
\usepackage{epsfig}
\usepackage{color}
\usepackage{subfigure}

\newtheorem{theorem}{Theorem}[section]

\newtheorem{proposition}[theorem]{Proposition}

\newtheorem{remark}[theorem]{Remark}

\def \beq{\begin{equation}}
\def \eeq{\end{equation}}

\def \ep {\epsilon}

\begin{document}

\begin{frontmatter}

\title{Smoothed dynamics in the central field problem}

\author{Manuele Santoprete}
\ead{msantoprete@wlu.ca}
\address{Department of Mathematics,
Wilfrid Laurier University,
Waterloo, N2L 3C5, Ontario, Canada}

\author{Cristina Stoica \corauthref{cor1}}
\ead{cstoica@wlu.ca}
\corauth[cor1]{corresponding author}

\address{Department of Mathematics,
Wilfrid Laurier University,
Waterloo, N2L 3C5, Ontario, Canada}

\begin{abstract} Consider the motion of a material point of unit mass in a central field determined by a  homogeneous potential of the form $(-1/r^{\alpha})$, $\alpha>0,$ where $r$ being the distance to the centre of the field.
Due to the singularity at $r=0,$ in computer-based simulations, usually, the potential is replaced by a similar potential that is  smooth, or at least continuous.

In this paper,  we compare the global flows given by the smoothed and non-smoothed  potentials. It is shown that the two flows are topologically equivalent for $\alpha < 2,$ while for $\alpha \geq 2,$  smoothing introduces fake orbits. Further, we argue that for  $\alpha\geq 2,$  smoothing should be applied to the amended potential $c/(2r^2)-1/r^{\alpha},$ where $c$  denotes the angular momentum constant. 


 \end{abstract}

\end{frontmatter}

\section{Introduction}

 For large particle systems, a principal tool of investigation is  computer-based simulation. In a variety of problems the interaction of the particles  is determined by  a  potential that is undefined at collisions. A common technique in dealing with the vector field singularities  is to replace the potential with a smooth, or at least continuous, function. This procedure is called smoothing, or, in physics terminology, softening.

Smoothing  was introduced in 1963 by S.J Aarseth cf. \cite{aaseth-phd}, \cite{aaseth}, 
in the context of numerical simulations of  galaxies. 
 Since then, smoothing has became a commonly used technique  in numerical modeling  of   large particle systems   (see for instance, \cite{dyer}, \cite{merritt}, \cite{MP03}  or \cite{neunzert} ).

Understanding the modifications induced by smoothing in large particle systems still remains a challenging task. A first step is to  look at systems formed by two particles, but even in this simplified context, one is faced with difficulties; see, for instance, the analysis presented in \cite{degiorgi}, where several  conjectures concerning  the convergence of the approximation methods are stated.

Closely related to smoothing is the concept of regularization: they both target singularities in the flow  as induced by the singularities in the   vector field, but the resolution is different. Smoothing modifies the vector field. Regularization  relies on a qualitative  analysis of the phenomena near singularity and is achieved in two distinct steps. First,  new parametrizations are applied, both time-dependent and -independent, leading  to a \textit{regularized vector field}, that is a vector field free of singularities. The phase space in the new coordinates is extended to include the singularity set, now blown-up into a physically fictitious and invariant manifold, usually called \textit{the collision manifold}. Second,  analysis of the flow on the extended phase space is performed in order to decide whether solutions asymptotically reaching the collision manifold can be matched to solutions asymptotically leaving the collision manifold, while preserving good behavior with respect to initial data. If such a matching is possible, then the flow may be   extended (at least continuously) to include orbits ending/starting in collision. When this  extension is performed, then the problem is said to be regularized. 
(For more on regularization, see  \cite{mcgehee} or, from a more physical point of view, see \cite{stoica}.) 

We also mention the paper of Bellenttini et al.  \cite{bellettini}, where  regularization is seen from the different perspective of approximating collision solutions by solutions of the smoothed flow. While analyzing a system where the interaction is given by homogeneous  potentials of the form $-{1}/{r^\alpha},$  $\alpha>0,$ the authors convey that their procedure leads to a larger set of  regularizable  problems than in the standard treatment.  Moreover, the smoothing chosen is \textit{irrelevant}, as long as it provides a flow free of singularities.

In this paper we question the appropriateness of smoothing when motion 
both near and \textit{far} from collision  is under scrutiny. Our analysis is performed within the class of homogeneous potentials to which a standard  potential smoothing \begin{equation} \label{smoothing-intro}U_{\epsilon}(r; \alpha):=-{1}/(r^2+\epsilon^2)^{\alpha/2}, \hspace{0.3cm}\epsilon >0,\end{equation} is applied. Within negative energy levels,  we focus on the topological equivalence of the non-smoothed and the smoothed flows outside the collision set.    We show  that for $\alpha \geq 2$ the two flows are \textit{not} topologically equivalent and thus smoothing of the form (\ref{smoothing-intro}) generates orbits that do not correspond to orbits of the real non-smoothed motion. For this case, we introduce the idea of smoothing the amended potential and show that, with such a modification, the two flows are topologically equivalent.

Employing a technique similar to that of McGehee \cite{mcgehee}, we choose to describe the dynamics  in a parametrization where the non-smoothed flow is nonsingular and where the phase space is extended to include the collision set, now blown-up into a one-dimensional manifold
(this is the first step of regularization as described above).  
 The orbits lie on compact three dimensional manifolds 
 which are level sets of the energy integral for negative energy values.
 Since the regularized vector field preserves the $SO(2)$ equivariance of the original problem, dynamics can be studied in a reduced three dimensional space.  Here orbits can be easily visualized as curves determined by the intersection of the two surface integrals, the energy and the angular momentum. This allows us to  compare  the orbital pictures of the non-smoothed versus the smoothed problem.

 While the non-smoothed reparametrized flow includes the orbits on the collision manifold, our analysis refers only to  orbits outside of it and concerns only orbit topology. We do not refer to regularization of solutions (the second step of regularization, as outlined above) and we do not focus on issues related to approximating the non-smoothed solutions by smoothed ones. 

 The paper is organized as follows: we begin by briefly reviewing known facts about dynamics of two particle systems. Next, we reparametrize the vector fields of the non-smoothed and smoothed problems such that the collision set of the non-smoothed problem is blown-up into the aforementioned collision manifold. Using the $SO(2)$ symmetry to reduce  the phase space to three dimensions, we study  relative equilibria and examine symmetries  of the reduced flow. Further, we analyse and compare of the orbits of the non-smoothed and smoothed flows, drawing the conclusion that for $\alpha\geq2$ the two flows are not topologically equivalent. In the last section we argue that for $\alpha\geq2,$ smoothing should be applied to both the potential and the rotational non-inertial term, leading to the idea of a smoothed amended potential. Moreover, we show that when such a smoothing is applied, the topological equivalence of the non-smoothed and smoothed flows outside the collision set is achieved.



\section{Equations of Motion} 

Consider the two degree of freedom Hamiltonian system given by the system of first order ordinary differential equations: 
\begin{equation}
\left\{ \begin{array}{l}
\dot{\bf q}=\frac{\partial H_{\epsilon}}{\partial {\bf p}},  \\
\dot {\bf p}=-\frac{\partial H_{\epsilon}}{\partial {\bf q}},\\
\end{array}
\right. 
\label{eqmotion}
\end{equation}
where ${\bf q}=(x,y)\in \mathbb{R}^2$, ${\bf p}=(p_x,p_y)\in \mathbb{R}^2$. The function \[H_{\epsilon}({\bf q}, {\bf p}):=\frac{ {\bf p}^2} {2}+ U_{\ep}(\bf{q};\alpha)\] is the Hamiltonian  of the system and  $U_\ep$ is a ``smoothed" potential 
\[
U_{\epsilon}({\bf q};\alpha)=-\frac{1}{(r^2+\epsilon^2)^{\alpha/2}}.
\]
where $\ep\geq 0$ is a parameter and $r= |{\bf q}|=\sqrt{x^2+y^2}$. For $\ep=0$ the potential $U_0({\bf q};\alpha)$ reduces to the classical homogeneous potential, in which case  the vector field defined by (\ref{eqmotion}) has a singularity at ${\bf q}={\bf 0}$. 

Since the system is Hamiltonian, it is well known that the total energy  is conserved.  
Consequently, the level sets of $H_{\epsilon}$ are invariant under the flow of 
(\ref{eqmotion})
 \[H_{\epsilon}({\bf q}(t), {\bf p}(t))=h=const.\] 
Due to radial symmetry,  the angular momentum is conserved as well and we have:  \[x(t) p_y(t)-y(t) p_x(t) =c=const.\]  
Therefore the system has two independent first integrals in involution and it is integrable by the Liouville-Arnold theorem.

Since $U_\epsilon({{\bf q},\alpha)}:\mathbb{R}^2 \setminus \{\bf{0}\} \rightarrow \mathbb{R}$ is real analytic, standard results of differential equation theory guarantee, for any initial data $({\bf q}_0,{\bf p}_0)=({\bf q}(0),{\bf p}(0))\in\mathbb{R}^2\setminus \{\bf{0}\}\rightarrow \mathbb{R} $, the existence and uniqueness of an analytic solution defined on a maximal interval $[0,t^*)$, where $0<t^*\leq \infty$. If $t^*<\infty$, the solution is said to experience a singularity. 

For $\ep=0,$ the singularity in the vector field induces singularities in the solution.  Singularities  specific to particle systems are given by  {\it collisions}, which occur when $\bf{q}(t)\rightarrow \bf{0}$  as $t \rightarrow t^*$.  In  \cite{mcgehee}, McGehee showed  that collisions are the only possible singularities of (\ref{eqmotion}). He also proved that if $\alpha\in (0,2),$ then $({\bf q}_0,{\bf p}_0)$ leads to a collision if and only if  the angular momentum is zero (that is ${\bf q}_0\times{\bf p}_0=0$) and that if $\alpha \geq 2,$ then the set of initial conditions leading to a collision is rather large, including for $\alpha>2$ zones  where the energy integral is negative. 

The standard methodology in dealing with collisions in n-particle simulations is  $smoothing$ the potential by setting $\ep>0.$ Then the vector field  (\ref{eqmotion}) is real analytic for all  initial data $({\bf q}_0,{\bf p}_0) \in \mathbb{R}^2 \times  \mathbb{R}^2,$ and the associated system of differential equations admits a unique global analytic solution.


\section{The flow of the smoothed potential}
\subsection{Topological description of the energy surfaces}

From now on, unless otherwise stated, the energy  $h$ is assumed to take negative values. 
 
Using a technique similar to McGehee \cite{mcgehee} we consider the following transformations $(r, p_r, \theta,  p_{\theta}) \to (r, v, \theta, u)$ defined by 
\begin{equation} 
\left\{\begin{array}{l}
u=p_{\theta} r^{\frac{\alpha-2}{2}},\\
v=p_{r} r^{\frac{\alpha}{2}}.\\
\end{array}
\right. \label{McGehee}
\end{equation}
This transformation is a diffeomorphism from  $(0,\infty) \times  \mathbb{R} \times \mathbb{S}^1 \times \mathbb{R}$ to itself (where $\mathbb{S}^1$ is the unit circle).  
Further, we rescale the time parametrization by 
\begin{equation} \label{time-para}
d\tau=r^{-\frac{\alpha+2}{2}}dt.
\end{equation}
It is useful to keep in mind $v$ and $u$ are re-parametrised  linear  and angular momenta, respectively. The equations of motion take the form
\beq\left \{\begin{array}{l}
r'=rv\\
v'=u^2+\frac{\alpha}{2}v^2-\frac{\alpha r^{\alpha+2}}{(r^2+\ep^2)^{\frac{\alpha}{2}+1}}\\
\theta'=u\\
u'=(\frac{\alpha-2}{2})uv,
\end{array}
\right.
\label{eqmcgehee}
\eeq
where prime denotes differentiation with respect to the independent variable $\tau.$ 
In the new coordinates the conservation of energy integral reads
\begin{equation}\label{energy-relation}
u^2+v^2-\frac{2r^\alpha}{(r^2+\ep^2)^{\alpha/2}}+2r^\alpha |h|=0.
\end{equation}
From above we deduce that  the phase space is foliated by the {\it energy surfaces}:  
\[ \Sigma_{h,\ep}:=\{ (r, v, \theta, u) \in [0,\infty)\times \mathbb{R}\times \mathbb{S}^1\times\mathbb{R} \hspace{0.1cm} | \hspace{0.1cm}  F_{\epsilon}(r,v,\theta,u)=0 \}, \] where 
\[
F_{\epsilon}(r,v,\theta,u):=u^2+v^2-\frac{2r^\alpha}{(r^2+\ep^2)^{\alpha/2}}+2r^\alpha |h|.
\]

\bigskip
 For $\epsilon =0,$ the transformations (\ref{McGehee}) and (\ref{time-para})  have  important consequences. First, the dynamics given by  (\ref{eqmcgehee}) has no singularity at $r=0$; in fact in the new coordinates  the system extends analytically to all of $(r,v,\theta,u)\in [0,\infty)\times \mathbb{R}\times \mathbb{S}^1\times\mathbb{R},$ and hence the equations of motion are regularized. Second, the submanifold $r=0$ is now invariant under the flow.  Third, the energy relation  also extends to the set $r=0$ giving 
\[u^2+v^2-2=0.\]
Let  
\beq  M_0:= \{ (r, v, \theta, u) \hspace{0.1cm} | \hspace{0.1cm}  r=0,  \hspace{0.2cm}  \theta \in \mathbb S^1,   \hspace{0.2cm}  u^2+v^2-2=0
 \},
\eeq that is $M_0$ is the boundary of the extended phase space. Note that, since  each energy surface has the same boundary $r=0,$ the set $M_0$ is independent of $h.$  Topologically,  it  is a two-dimensional manifold embedded in $\mathbb R^3\times\mathbb S^1$ and it is diffeomorphic to a torus. We call   $M_{0}$   the {\itshape collision-ejection} manifold or, simply, the {\itshape collision manifold}.

In the new parametrization, orbits which previously reached  $r=0$ in a finite time  now tend asymptotically toward collision manifold. Further, orbits which previously passed close to collision now spend a long time  near $M_0$.
By continuity with respect to initial data, the flow on collision manifold, although lacking physical meaning, provides useful information about collision and near-collision solutions.

\bigskip
For $\ep>0$, using the energy relation (\ref{energy-relation}), we deduce that  on each $\Sigma_{h,\ep}$ the radial coordinate is bounded by:
\beq \label{r-max}
0\leq r\leq  R^{max}_{\ep}=\sqrt{\left(\frac{1}{|h|}\right)^{2/\alpha}-\ep^2}.
\eeq
 In physical space this means that the maximal distance in between points is bounded by the value $R^{max}_{\ep}$. 

\medskip
 \begin{remark}
Inequality (\ref{r-max}) is meaningful provided $\ep$ is small enough, that is, if \[\ep < \left(\frac{1}{|h|} \right)^{\frac 1 \alpha}.\] Henceforth we will assume this condition is always satisfied. 

 \end{remark}

 \bigskip
 The energy surface $\Sigma_{h,0}$ is  a smooth manifold.  For  $\ep>0,$ if  $\alpha\leq 1,$ $\Sigma_{h,\ep}$  is also a smooth manifold;   if  $\alpha>1,$ the energy surface  is not smooth at $(0,0,\theta,0)$ since   $\nabla F_{\ep}$ at $(0,0,\theta,0)=(0,0,0,0)$ as it can be readily seen from 
\[
\frac{dF_{\epsilon}}{dr}=r^{\alpha-1}\left(\frac{-2\alpha}{(r^2+\ep^2)^{\alpha/2}}+\frac{2r^2\alpha}{(r^2+\ep^2)^{\alpha/2+1}}+2\alpha |h|\right).
\] 

\bigskip
\begin{proposition}
For  $\ep>0$ and negative energy $\Sigma_{h, \ep}$ is diffeomorphic to $\mathbb{S}^2\times \mathbb{S}^1$ for $\alpha \leq 1$ and homeomorphic to it for $\alpha> 1$.
\end{proposition}

\smallskip
\noindent {\textbf {Proof}}
In the  space  $(r,u,v)\in [0,\infty)\times\mathbb{R}^2,$ $H_{\epsilon}(r,v,\theta,u)=0$ describes a surface of revolution. This surface is orientable, compact, connected and  of genus $0$. It is  smooth for $\alpha\leq 1$, and thus it is diffeomorphic to a two-sphere $\mathbb{S}^2$. It is not smooth for $\alpha>1$ and therefore it is homeomorphic to a sphere.  Consequently, for $\alpha\leq 1$ ($\alpha>1$), $\Sigma_{h, \ep}$ is diffeomorphic (homeomorphic) to  $\mathbb{S}^2\times \mathbb{S}^1$. 
$\square$

\bigskip
In the new coordinates, the conservation of the angular momentum translates into the presence of invariant surfaces of the form
\begin{equation}\label{ang-mom}
\Gamma_c=
\begin{cases}
   \{(r, v, \theta, u) \in [0,\infty)\times \mathbb{R}\times \mathbb{S}^1\times\mathbb{R} \hspace{0.1cm} | \hspace{0.1cm} u r^{\frac{2-\alpha}{2}}=c\}   & \text{if} \quad \alpha \leq 2 \\
   \{(r, v, \theta, u) \in [0,\infty)\times \mathbb{R}\times \mathbb{S}^1\times\mathbb{R} \hspace{0.1cm} | \hspace{0.1cm} u =cr^{\frac{\alpha-2}{2}}\}   & \text{if} \quad \alpha > 2.
\end{cases}
\end{equation}
Taking into account energy conservations, we deduce that the phase space is foliated by invariants sets obtained as intersections  $\Sigma_{h,\ep} \cap \Gamma_c.$ 

\subsection{Reduced dynamics}

We now return to the system of equations  (\ref{eqmcgehee}). In order to study the dynamics  it is convenient to exploit the fact that $\theta$ does not appear explicitly in the equations. This allows us to reduce the four-dimensional phase space  by factoring out the flow by $ \mathbb{S}^1.$ The three dimensional reduced phase space is described by $(r,v,u) \in [0, \infty) \times \mathbb{R} \times  \mathbb{R}$  and a vector field given by: 
\beq\left \{\begin{array}{l}
r'=rv\\
v'=u^2+\frac{\alpha}{2}v^2-\frac{\alpha r^{\alpha+2}}{(r^2+\ep^2)^{\frac{\alpha}{2}+1}}\\
u'=(\frac{\alpha-2}{2})uv.
\end{array}
\right.\label{red-eqmcgehee}
\eeq Defining  \[f_{\ep}(r):=\frac{2r^\alpha}{(r^2+\ep^2)^{\alpha/2}}-2r^\alpha |h|,\] the energy relation (\ref{energy-relation}) reads:
\begin{equation}\label{eq:energy-relation}
u^2+v^2=f_{\ep}(r).
\end{equation}
With a slight abuse in notation, we choose to call the energy surfaces in the reduced space 
\begin{equation} \label{energy-red} \Sigma_{h, \ep}:= \{ (r, v,u) \in[ 0,\infty)\times \mathbb{R} \times\mathbb{R} \hspace{0.1cm}| \hspace{0.1cm}   u^2+v^2-f_{\ep}(r)=0 \}. \end{equation} Likewise,  the angular momentum invariant sets given by (\ref{ang-mom}) are: \begin{equation} \label{ang-mom-red} 
\Gamma_c=
\begin{cases}
   \{(r, v,  u) \in [0,\infty)\times \mathbb{R}\times \mathbb{R} \hspace{0.1cm} | \hspace{0.1cm} u r^{\frac{2-\alpha}{2}}=c\}   & \text{if} \quad \alpha \leq 2 \\
   \{(r, v,  u) \in [0,\infty)\times \mathbb{R}\times \mathbb{R} \hspace{0.1cm} | \hspace{0.1cm} u =cr^{\frac{\alpha-2}{2}}\}   & \text{if} \quad \alpha > 2.
\end{cases}
\end{equation}

\subsection{Relative equilibria}
We now study the existence and nature of the equilibria of the reduced dynamics. 
Such fixed points which, in addition, are located outside the collision manifold (that is, those with $r>0$)  correspond in the unreduced space to relative equilibria. In physical space,  they represent  uniform circular motions around the center of mass. 

\medskip
\noindent {\textbf {Case $\ep=0$}}.  Fixed points with $r=0$ are found on the collision manifold $M_0$ and are situated at $C_{\pm}= (0, \pm \sqrt{2}, 0).$ Direct calculations show that they are saddles. 
In physical space they correspond to {\it radial} fall to or ejection from $M_0.$ The radial coordinate of fixed points with $r>0$ is given by:
\[
2|h|r^\alpha=(2-\alpha).
\] 
This equation has a unique solution 
\[
r_e=\left (\frac{2-\alpha}{2|h|}\right)^{\frac 1 \alpha}
\]for $\alpha<2$ and no positive  solutions if $\alpha\geq 2.$
Thus relative equilibria are present only for  $\alpha<2$ and they are of the from 
$R_0^{\pm}=(r_e,0,\pm u_e),$ where $u_e$ may be determined by substituting $r=r_e$ and $v=0$ into the energy relation (\ref{eq:energy-relation}). A  direct verification shows  that they are centres. 

This is not surprising, since it is well-known that motion in an attractive potential of the form $1/r^{\alpha},$ $\alpha<2$  possess such equilibria in the reduced space (corresponding to circular orbits in the physical space) as long as the angular momentum is non-zero (see \cite{golstein}).

\vspace{0.3cm}
\noindent {\textbf {Case $\ep>0$}} 
It is immediate that the origin $O:=(0, 0,0)$ is a fixed point. The remaining solutions,  if they exist, are of the form $R_\ep^{\pm}=(r_{\ep},0, \pm u_\ep).$ Setting $v=0$ in the second  equation of (\ref{red-eqmcgehee}), we obtain
\[
u^2-\alpha\frac{r^{\alpha+2}}{(r^2+\ep^2)^{\frac \alpha 2+1} }=0
\]
and, since by the energy relation
\[u^2-f_{\ep}(r)=u^2 - \frac{2r^\alpha}{(r^2+\ep^2)^{\frac{\alpha}{2} }} +2r^\alpha |h|=0,\] 
 we have to solve 
\[-\alpha\frac{r^{\alpha+2}}{(r^2+\ep^2)^{\frac \alpha 2+1} }= - \frac{2r^\alpha}{(r^2+\ep^2)^{\frac \alpha 2}}+2r^\alpha |h|.\]
 This amounts to studying the zeroes of the 
function
$
r^{\alpha}/(r^2+\ep^2)^{(\alpha/ 2 +1)}h(r)$
where \[h(r):=(2-\alpha)r^2+2\ep^2-2|h|(r^2+\ep^2)^{\frac \alpha 2 +1}.\]
 If $\ep $ is sufficiently small,  $h(0)=2\ep^2(1-|h|\ep^\alpha)>0$ and $h(r)\rightarrow -\infty$ as $r\rightarrow \infty$. 
This implies that $h(r)$ has at least one zero for $r>0$.
The derivative is $h'(r)=2rg(r)$ with 
\[g(r):= (2-\alpha)-2|h|(\frac \alpha 2 +1)(r^2+\ep^2)^{\frac \alpha 2}.\]
If $0<\alpha<2,$ $g(r)$ is positive for $r<r^*$ and negative for $r>r^*$
where \[r^*=\sqrt{\left(\frac{2-\alpha}{2|h|(\frac \alpha 2 +1)}\right)^{\frac 2 \alpha}-\ep^2}.\]
Consequently, $h(r)$ increases for $r<r^*$ and decreases for $r>r^*,$  and since $h(0)>0,$ the equation $h(r)=0$ has an unique solution. If $\alpha\geq 2,$  $g(r)<0$ thus $h'(r)<0$,  $h(r)$ is a monotone decreasing function and $h(r)=0$ has a unique positive solution.

Hence, for any $\alpha>0$ the reduced smoothed flow admits two  relative equilibria located at $R_\ep^{\pm}=(r_{\ep},0, \pm u_\ep).$

\subsection{Symmetries}

Observe that the plane $u=0$ is an invariant manifold. The dynamics is symmetric with respect to  transformations of the form:
\[
S~:~~ (r,v,u,\tau)\rightarrow (r,-v,u,-\tau).
\]
The invariance under this symmetry implies that if $\gamma(\tau)$ is a solution of (\ref{eqmcgehee}) then also $S(\gamma(\tau))$ is a solution. We have the following:

 \medskip
\begin{proposition} \label{symmetry}
If an orbit  crosses the plane $v=0$ in one point it is $S-$symmetric.
If it crosses the plane $v=0$ in two (distinct) points (with $r>0$) it is $S-$symmetric and periodic. \label{lemmasymm}
\end{proposition}

\smallskip
\noindent {\textbf {Proof}}
The first part of the theorem follows from the uniqueness of solutions. The second part follows since the solution is $S-$symmetric and closed (and does not intersect a critical point).
$\square$
\begin{figure}[t] 
\centerline{\includegraphics[angle=0, scale=.8]{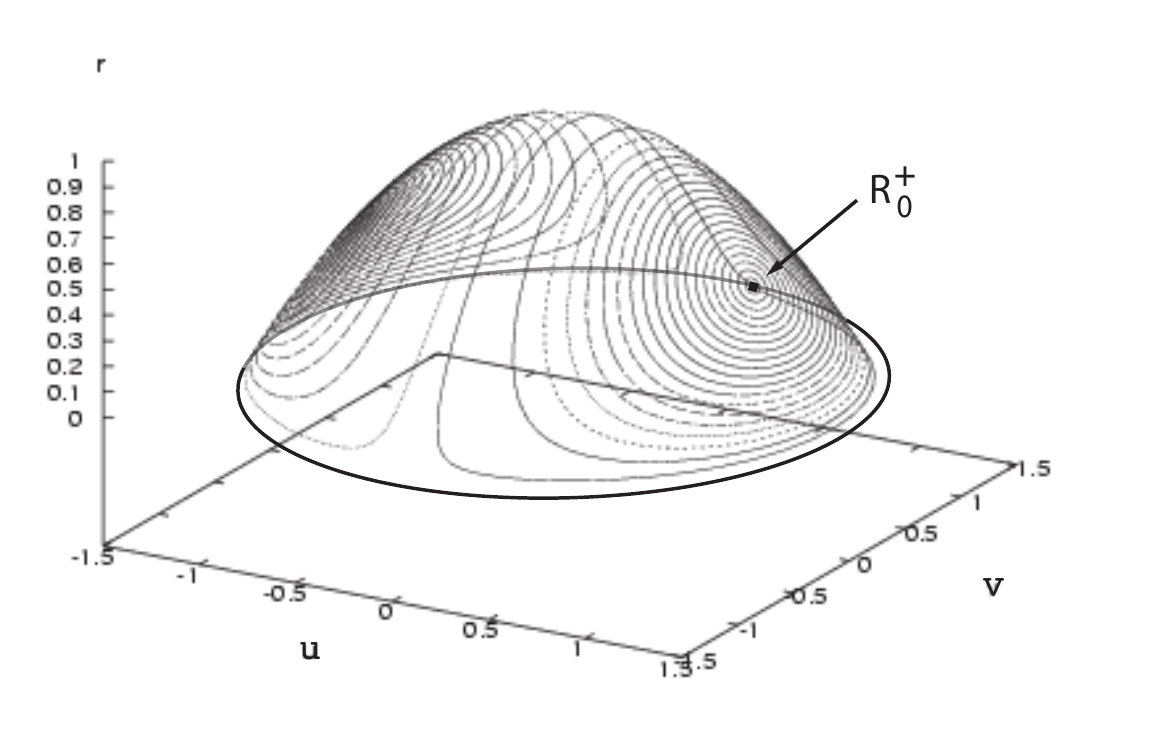}}
\caption{The reduced orbit space for  $\ep=0$ and $\alpha=1$ outside the collision manifold, i.e. on the domain $r>0.$  The energy surface $\Sigma_{h,0}$ is of a shape of a hemisphere. The orbits are given by the intersections of $\Sigma_{h,0}$ with the angular momentum surfaces (not represented here). Easily noticed, in this case all orbits are periodic.}
\end{figure}

\bigskip
Recall that for each fixed level $h<0$ of the energy, the reduced flow is constrained  to the two dimensional surface $\Sigma_{h, \ep}$. Since the angular momentum is conserved, the orbits are  determined by the intersection of $\Sigma_{h, \ep}$ with the two-dimensional angular momentum surfaces $\Gamma_c$ (see Figure 1).  In other words, having $\ep \geq 0$ fixed, to each pair  energy-angular momentum pair $(h,c)$ it corresponds a unique orbit 
\begin{align*} \gamma^{h,c}_{\ep}& :=  \Sigma_{h, \ep} \cap \Gamma_c. 
\end{align*}

  Using Proposition  \ref{symmetry}, it is sufficient to study the intersection of each orbit with the plane $v=0.$ Note that,  on a given orbit, points of the form $(r,0,u)$ with $r>0$ are turning points.  Thus, for a periodic orbit, the two cuts  with $v=0$ are the points where the relative distance $r$ in between particle attains its maximum  and its minimum.  For a non-periodic orbit, the cut $v=0$ corresponds to the  maximal relative distance.

Since the graph of  $\Sigma_{h, \ep}$ is symmetric with respect to the horizontal axis $u=0,$ we can restrict our domain to $u\geq0.$ Note that from (\ref{ang-mom-red}) we have that $u\geq 0$ if and only if  $c\geq 0.$ 
So, fixing $h<0,$  the orbits are given by  the intersection of 
\[
u_{h,\ep}(r):= + \sqrt{f_{\ep}(r)},\hspace{0.3cm}\ep\geq0,\] with the curves $u_c(r)$ as defined by 
\[
 \begin{cases} 
 u_c(r)r^{\frac{2-\alpha}{2}}=c
     & \text{if} \quad \alpha \leq 2, \quad c\geq 0,\\
 u_c(r) =cr^{\frac{\alpha-2}{2}}   & \text{if} \quad \alpha > 2, \quad c \geq 0.
\end{cases} 
\]
   If this intersection is void, then there is no orbit, whereas 
if the $u_{h,\ep}(r)$ and $u_c(r)$ intersect in one (non-tangential) point, then  the corresponding orbit is a fall to/escape from collision. 
Further, if the $u_{h,\ep}(r)$ and $u_c(r)$ have  two distinct points in common, then the corresponding orbit is periodic. Finally, if $u_{h,\ep}(r)$ and $u_c(r)$ are tangent to each other,  then the point of tangency corresponds to  a relative equilibrium. 

\subsection{Topological equivalence of the reduced flows}

Let $\ep>0$ and consider the intersections of $u_{h,0}(r)$ and $u_{h,\ep}(r)$ with $u_c(r),$ where $c\geq0.$



\begin{figure}[t] 
\centerline{\includegraphics[angle=0, scale=0.7]{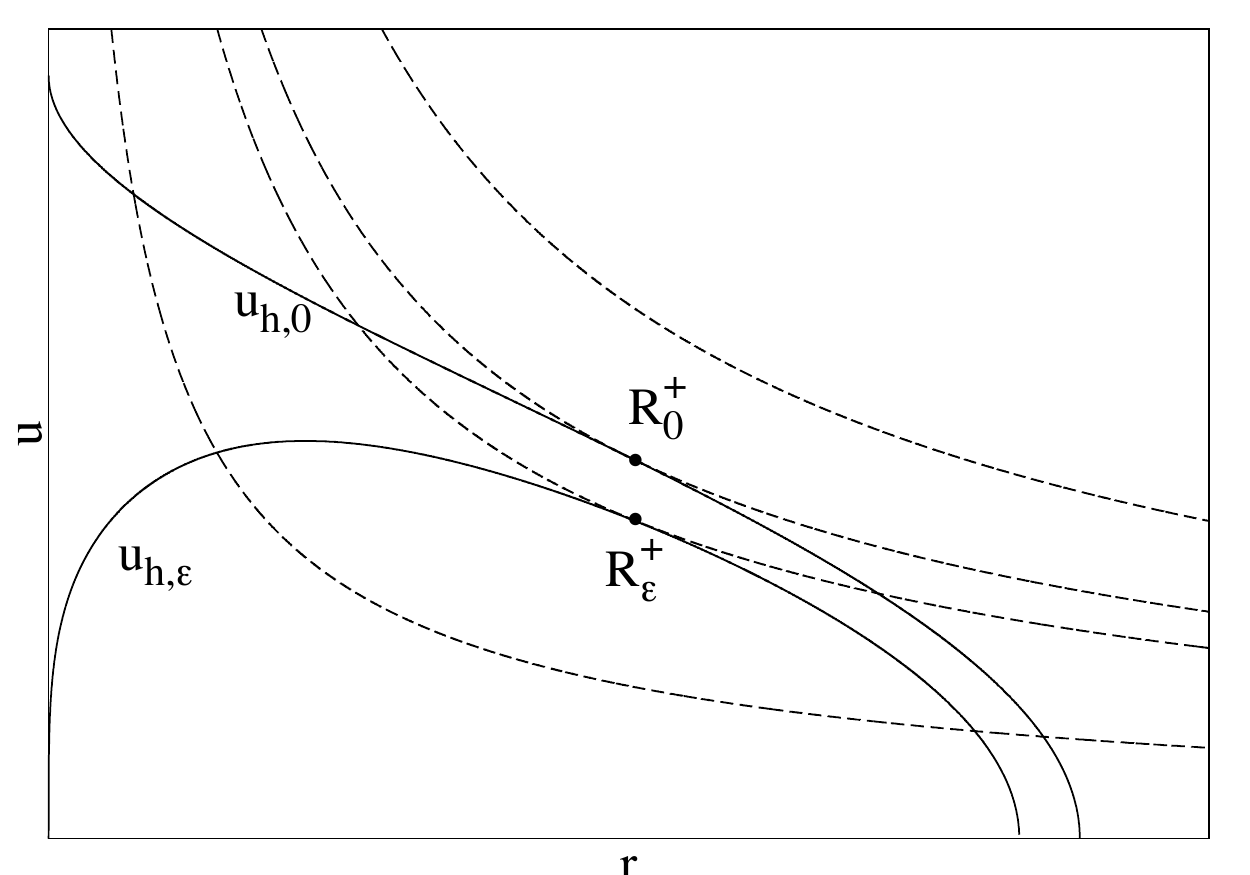}}
\caption{Intersections of $u_{h,0}(r)$ (top solid curve) and $u_{h,\ep}(r)$ (bottom solid curve) with  $u_{c}(r)$ (dashed curves). Case $\alpha<2.$}\label{flow-alpha<2}
\end{figure}

\medskip
\noindent{\textbf {Case $\alpha<2.$}} For $c>0,$ all orbits are periodic, exception being the fixed points $R_0^+$ and $R^{+}_{\ep}.$ For $c=0,$ all the orbits leading to/ejecting  from collision  (see Figure \ref{flow-alpha<2}). 
Note that  outside the collision set $c=0,$ the real and smoothed flows are topologically equivalent. 

\begin{figure}[t]
\centerline{\includegraphics[angle=0, scale=0.7]{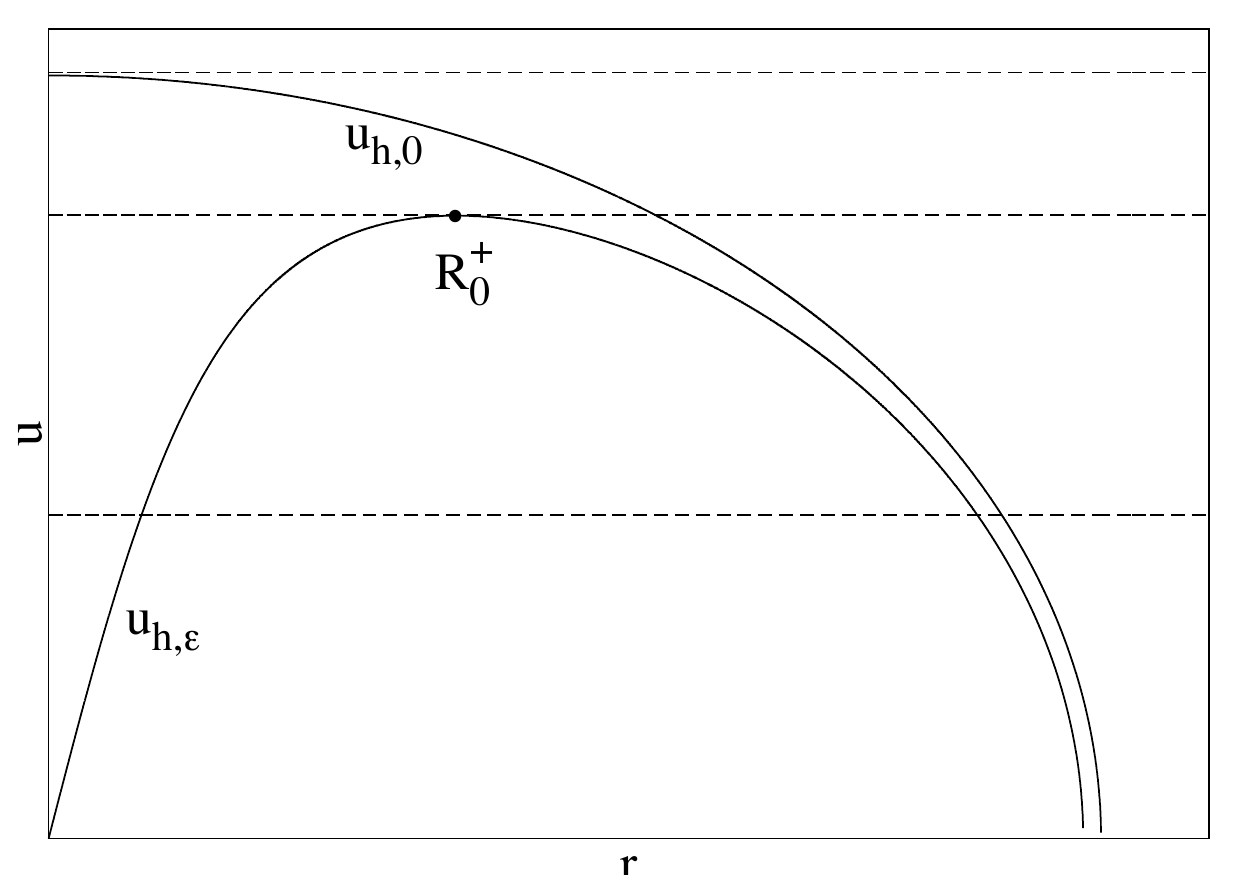}}
\caption{Intersections of $u_{h,0}(r)$ (top solid curve) and $u_{h,\ep}(r)$ (bottom solid curve) with  $u_{c}(r)$ (dashed curves). Case $\alpha=2.$
}\label{flow-alpha=2}
\end{figure}

\medskip
\noindent
{\textbf {Case $\alpha=2.$}} The angular momentum curves become horizontal lines $u_c(r)=c$  (see Figure \ref{flow-alpha=2}). All orbits of the real flow are either  leading to/escaping from collision, whereas all orbits of the smoothed flow are periodic. The real and smoothed flows are not topologically equivalent.

\begin{figure}[t]  
\centerline{\includegraphics[angle=0, scale=0.7]{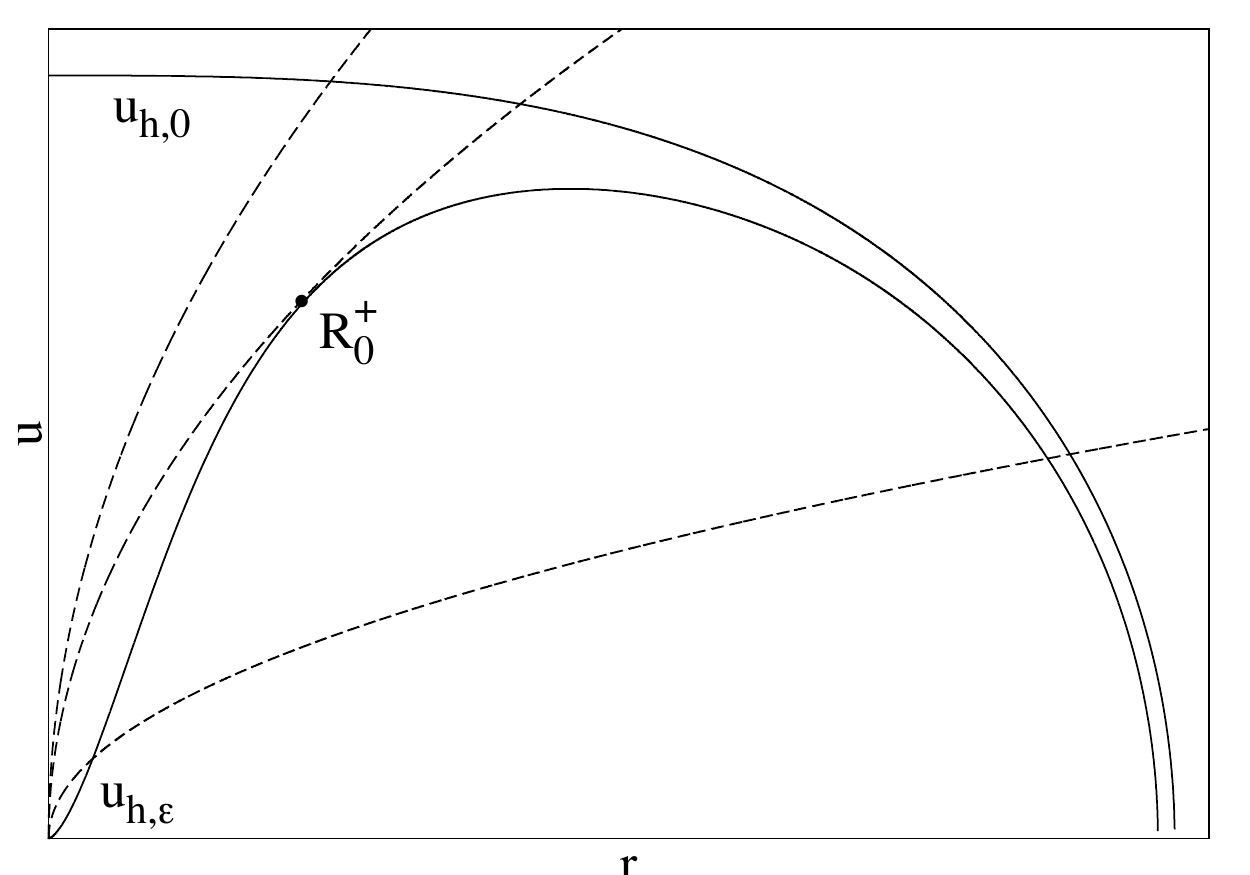}}
\caption{Intersections of $u_{h,0}(r)$ (top solid curve) and $u_{h,\ep}(r)$ (bottom solid curve) with  $u_{c}(r)$ (dashed curves). Case $\alpha>2.$ ($\alpha=3$). Mo
}\label{flow-alpha>2}
\end{figure}

\medskip
\noindent 
{\textbf {Case $\alpha>2.$}} First, note that all  orbits of the real flow are all  leading to/escaping from collision (see Figure \ref{flow-alpha>2}).

To determine the nature of the smoothed flow, we have to solve $u_{h,\epsilon}(r)=u_c(r).$ That is $\sqrt {f_{\epsilon}(r)}= cr^{\frac{\alpha-2}{2}},$ and further\beq
r^\alpha\left[\frac{c^2}{r^{2}}-\frac{2}{(r^2+\ep^2)^{\alpha/2}}-2 h\right]=0.
\label{eq}
\eeq

If $c=0,$ the previous relation becomes
\[\frac{1}{(r^2+\ep^2)^{\alpha/2}}=|h|.\] The positive solution is  unique  and it is given by $r_m:=\sqrt {(1/|h|)^{2/\alpha} -\epsilon^2}.$ This solution corresponds to an  ejection-collision orbit with no spin.

If $c\neq 0,$ let us consider $\chi(r):=\frac{c^2}{r^2}-\frac{2}{(r^2+\ep^2)^{\frac \alpha 2}}.$
 The positive solutions of $\chi(r)=0$ are given by the positive solutions of $\psi(r):=-\left(\frac{2}{c^2}\right)^{\frac 2 \alpha}r^{\frac 4 \alpha}+r^2+\ep^2=0$.
Note that $\psi(0)=\ep^2,$ and $\psi(r)\rightarrow \infty$ as $r\rightarrow \infty.$ Also,  
$\psi(r)$ has a unique minimum attained for $r=r_m:=\left(\frac{2}{\alpha}\left(\frac{2}{c^2}\right)^{2/\alpha}\right)^{\frac{\alpha}{2\alpha-4}}$. Moreover $\psi(r_m)=\left(\frac{2}{\alpha}\left(\frac{2}{c^2}\right)^{2/\alpha}\right)^{\frac{4}{2\alpha-4}}\left(\frac{2}{c^2}\right)^{2/\alpha}\left[2/\alpha-1\right]+\ep^2<0$ provided $\ep$ is sufficiently small. 
It follows that $\psi(r)$ has two zeroes and thus $\chi(r)$ has two zeroes.

Thus $\chi(r)-2h$ has two solutions when $h<0$ and $h$ is larger than the minimum of $\chi(r)$, one solution when  $h$ is equal to the minimum of $\chi(r)$ and zero solutions otherwise. We deduce that for $h$ in the domain of interest  (that is for non-void intersections of $u_{h,\epsilon}(r)$ with $u_c(r)$) and for $c\neq0,$ the orbits for the smoothed flow  are periodic, provided that $\epsilon$ is small enough. 
Since the real flow consists of only collision/ejection orbits,  the real and the smoothed flows are not topologically equivalent. 

\section{The flow of the smoothed amended potential}

In the previous section we have shown that for $\alpha \geq 2$ the flow associated to the smoothed potential is not topologically equivalent to the real flow. To understand why this is the case  we return to the initial set-up  of the problem. 
 
 Recall the Hamiltonian of the real flow \beq H_0(x,y, p_x, p_y)= \frac{1}{2} \left( p_x^2+ p_y^2  \right) - \frac{1}{\sqrt{(x^2+y^2)^{\alpha}}}\eeq or in polar coordinates: \beq  \label{H-zero} H_0(r, \theta, p_r, p_{\theta})= \frac{1}{2}\left(p_r^2+ \frac{p_{\theta}^2}{r^2}\right)- \frac{1}{r^{\alpha}}.\eeq 
It is clear from the expression above that the behaviour near collision is dominated by the term $1/r^{max\{2,\alpha\}}.$ It follows that for $\alpha<2,$  modifying the potential {\it only} does not change the main features of the flow near collision. This is not the case for $\alpha \geq2.$ Here, smoothing of the dominant term allows  the centrifugal term ${p_{\theta}^2}/{r^2}$ to dominate  near collision and to change the character of the motion.

\bigskip
Recall 
 that for the motion of a point in a central field on the plane, the distance from the centre of the field varies as in the one dimensional problem with a potential 
 :\[V(r):= \frac{c^2}{2r^2}+ U(r), \] where $c:= p_\theta(t)=const.$ is the conserved angular momentum (see, for instance \cite{Arnold}). The function 
$V(r)$ is usually called the {\it amended} or {\it effective potential}. The regions of motion together with the orbits'  type  (i.e. bounded or unbounded) are determined by  the inequality $V(r)\leq h.$ 

\begin{figure}  \centerline{\includegraphics[angle=0, scale=0.7]{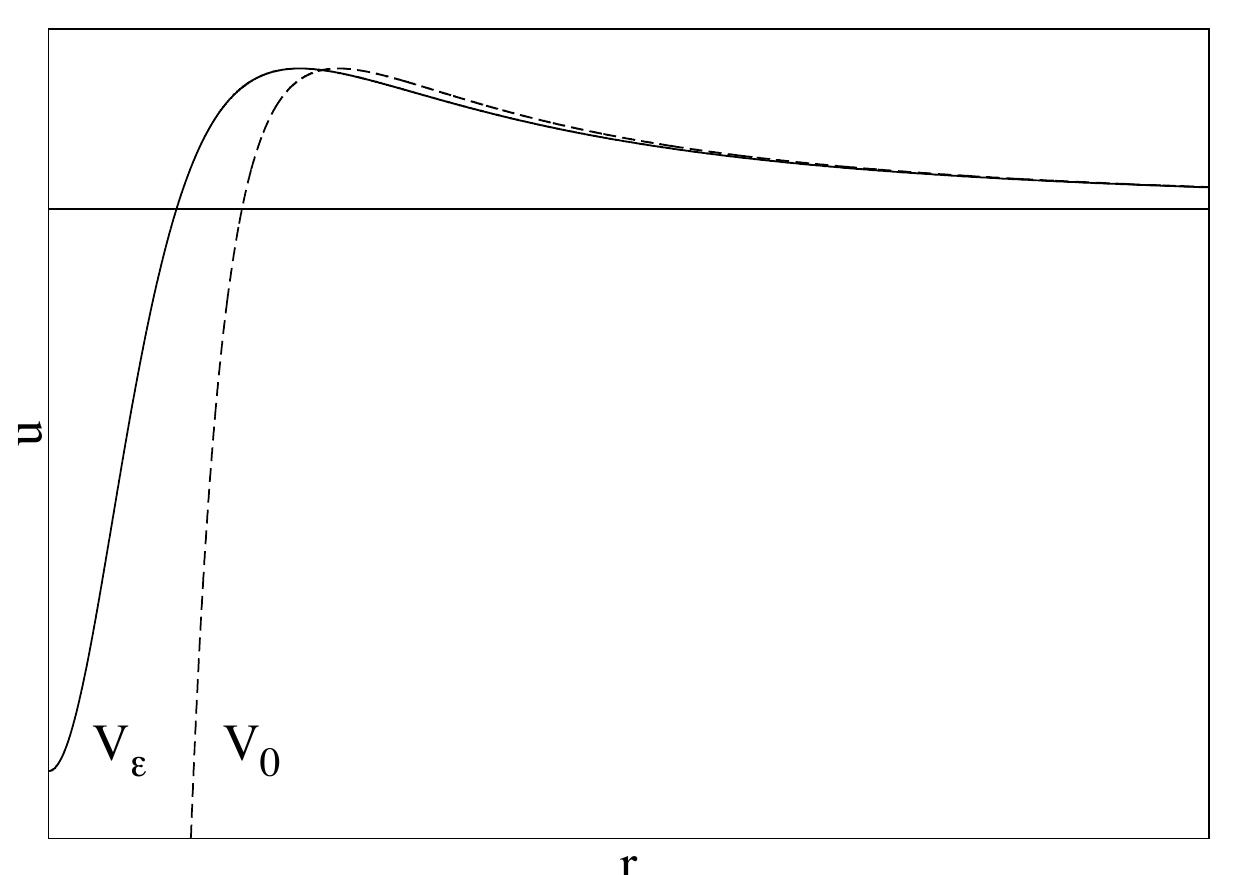}}
\caption{The non-smoothed amended  potential $V_{0}$ and  the smoothed amended  $V_{\ep}$ potential for $\alpha \geq2.$
}\label{V-eff}
\end{figure}

When smoothing is applied, the preservation, as much as possible, of the character of the orbits is necessary. For the homogeneous interaction with $\alpha \geq 2$, this may be  achieved by  smoothing the amended potential:
  \[V_{\ep}(r):= \frac{c^2}{2(r^2+\ep^2)}- \frac{1}{(r^2 +{\ep}^2)^{\alpha/2}}.\] 
 As  readily seen from Figure \ref{V-eff},  the orbits are of similar type in both non-smoothed and smoothed problems,  as long as  $c$  is chosen so that $V_{\ep}(0)<0.$ The regions of motion are determined by  $V_{\ep}(0)<h<0$ and so, to have a non-void orbit set,   $c$ must be chosen such that:
\beq \label{bound} |c| \leq \sqrt{\frac{2(1-|h|\ep^{\alpha})}{\ep^{\alpha-2}}}.\eeq

\bigskip
To decide the topological character or the orbits, we return to initial set-up  and consider  the smoothed Hamiltonian \beq H_{\ep}(r, \theta, p_r, p_{\theta}):= \frac{1}{2}\left(p_r^2+ \frac{p_{\theta}^2}{(r^2+\ep^2)}\right)- \frac{1}{(r^2 +{\ep}^2)^{\alpha/2}}, \quad \alpha \geq2. \eeq 
Now we  perform a study similar to  one in the previous section. Omitting the details, the main steps are:

\vspace{0.3cm}\noindent 
- first, we apply  transformations similar to (\ref{McGehee}) and (\ref{time-para}):
\begin{equation} \left\{\begin{array}{l}
u=p_{\theta} (r^2 +{\ep}^2)^{\frac{\alpha-2}{4}},\\
v=p_{r} (r^2 +{\ep}^2)^{\frac{\alpha}{4}};\\
\end{array}
\right. \label{new-McGehee}
\end{equation}
and introduce a new time parametrization via:
\begin{equation} \label{repara}
d\tau=r^{-\frac{\alpha+2}{2}}dt;
\end{equation}

\vspace{0.3cm}\noindent 
- the new vector field is regularized and is given by:
\beq\left \{\begin{array}{l}
r'=vr^{\frac{\alpha+2}{2}}(r^2 +{\ep}^2)^{-\frac{\alpha}{4}},\\
v'=r^{\frac{\alpha+4}{2}}(r^2 +{\ep}^2)^{-\frac{\alpha+4}{2}}(u^2+\frac{\alpha}{2}v^2-\alpha),\\
\theta'=ur^{\frac{\alpha+2}{4}}(r^2+\ep^2)^{-\frac{\alpha+2}{4}},\\
u'=(\frac{\alpha-2}{2})r^{\frac{\alpha+2}{4}}(r^2 +{\ep}^2)^{-\frac{\alpha+4}{4}}uv; 
\end{array}
\right.
\label{new-eqmcgehee}
\eeq 

\vspace{0.3cm}\noindent 
- since $\theta$ does not appear explicitly in the $(r,v,u)$ equations above, we reduce the phase space to three dimensions by factoring out the flow by ${\cal S}^1;$ 

\vspace{0.3cm}\noindent 
- in the reduced space $(r,v,u)$ the energy surfaces take the form
\[\Sigma_{h,\ep}= \{(r,v, u)  \hspace{0.1cm} |  \hspace{0.1cm} u^2 +v^2= 2\left(1-  |h|(r^2+ {\ep}^2)^{\alpha/2} \right)    \};  \] they are surfaces of revolution and, as it may be easily noticed, their topology is identical;

\vspace{0.3cm}\noindent 
- the angular momentum surfaces are given by

\[\Gamma_c=\{(r,v,u)\hspace{0.1cm} |  \hspace{0.1cm} u= c  (r^2 +{\ep}^2)^{\frac{\alpha-2}{4}} \};\] 

\vspace{0.3cm}\noindent 
-by symmetry, each orbits may be represented by its intersection with the plane $v=0.$ Therefore, it is sufficient to study the intersections of the curves \[u_{h,\ep}(r)=\sqrt{f_{\ep}(r)}:=\sqrt{2\left(1-  |h|(r^2+ {\ep}^2)^{\alpha/2} \right)},  \] with \[u_{c, \ep}(r):=c  (r^2 +{\ep}^2)^{\frac{\alpha-2}{4}}, \hspace{0.2cm} c\geq0.\]

\begin{figure}[t]
  \begin{center}
    \mbox{
      \subfigure[]{\resizebox{!}{4.2cm}{\includegraphics{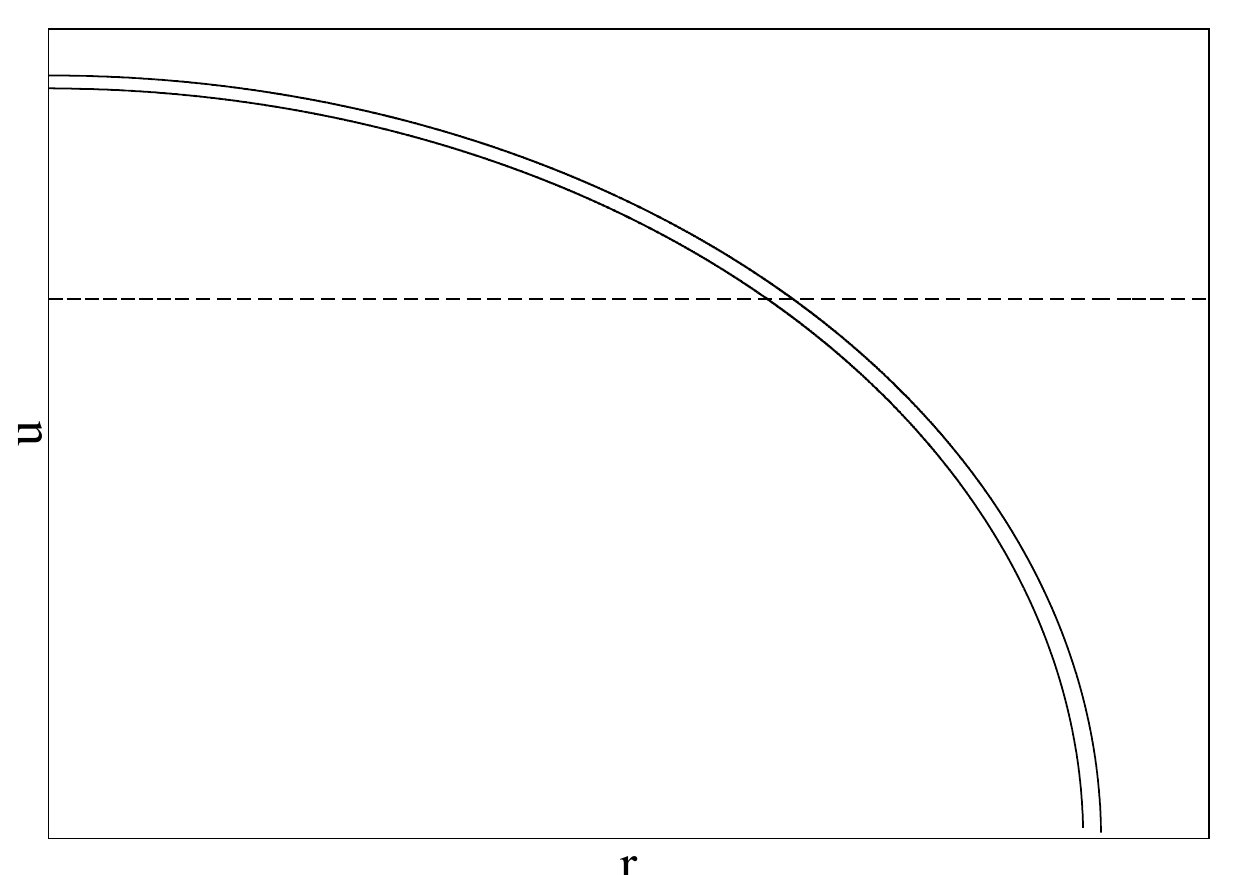}} \label{triangle}} \quad
      \subfigure[]{\resizebox{!}{4.2cm}{\includegraphics{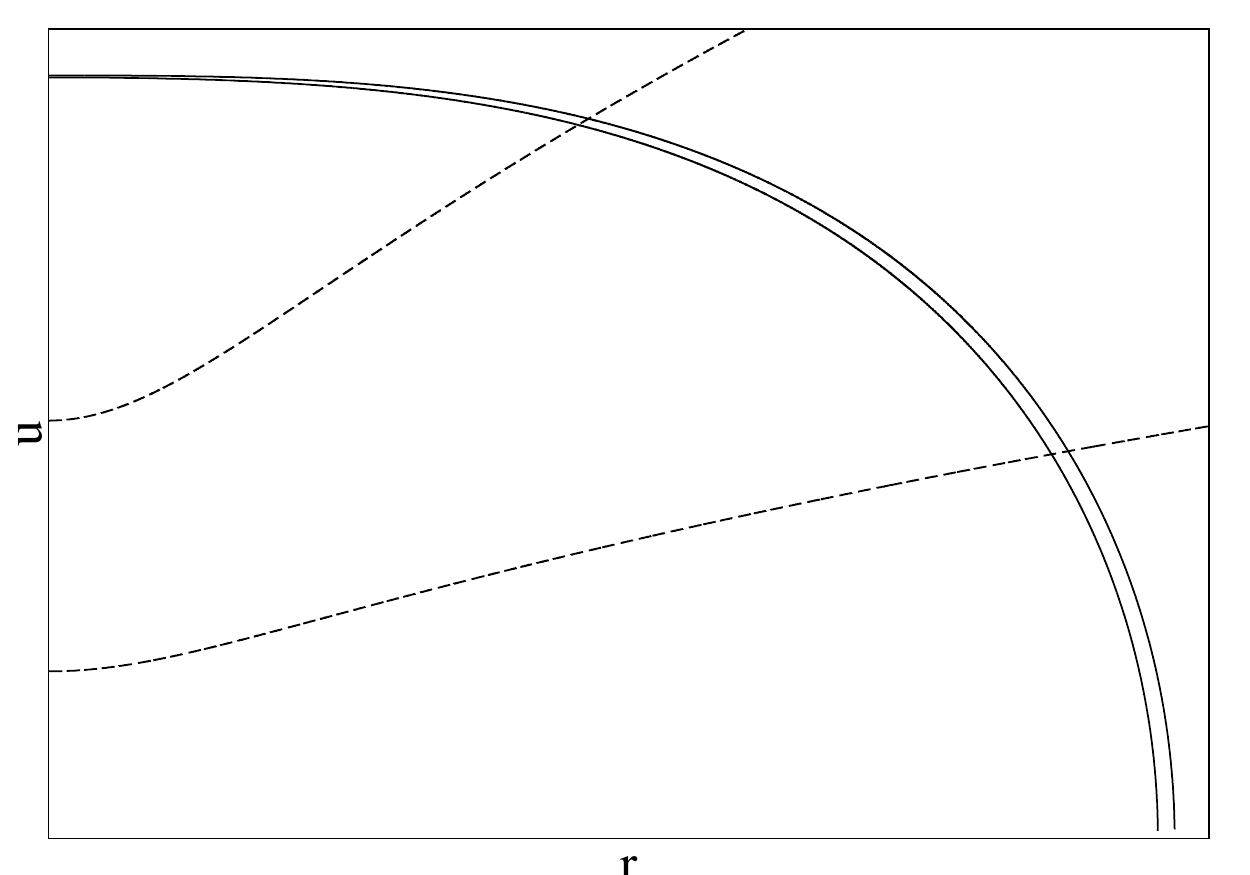}} \label{triangle1}}} 
    \caption{(a) Intersections of $u_{h,0}(r)$ (top solid curve) and $u_{h,\ep}(r)$ (bottom solid curve) with  $u_{c}(r)$ (dashed curves). Case $\alpha=2.$  (b)Intersections of $u_{h,0}(r)$ (top solid curve) and $u_{h,\ep}(r)$ (bottom solid curve) with  $u_{c}(r)$ (dashed curves). Case $\alpha>2.$ }
  \end{center}
\end{figure}

\bigskip
For $\alpha\geq 2$ all orbits are non-periodic, leading to or ejecting from the collision manifold $M_{\ep}:= \{(r,v, u)  \hspace{0.1cm} |  \hspace{0.1cm} r=0, \hspace{0.1cm} u^2 +v^2= 2\left(1-  |h|( {\ep}^2)^{\alpha/2} \right)    \} .$ For $\alpha=2,$ the angular momentum curves $u_{c, \ep}(r)$ are horizontal lines (see Figure 5), whereas for $\alpha>2,$ they are increasing and have initial value   $u_{c,\ep}(0)=c  {\ep}^{\frac{\alpha-2}{2}}$ (see Figure 6). The admissible values for $c$  are obtained by requiring $u_{c,\ep}(0)\leq  u_{h,\ep}(0).$ Thus we obtain $c  {\ep}^{\frac{\alpha-2}{2}} \leq \sqrt{2\left(1-  |h|{\ep}^{\alpha} \right)},$ which,  after some algebra, becomes condition (\ref{bound}). More importantly, the real and smoothed flows are topologically equivalent.

\section{Conclusions}
In this paper we discussed the topological equivalence of the non-smoothed and the smoothed flows given by motion with a homogeneous   potential of the form $-1/r^
{\alpha},$ $\alpha >0.$  The analysis was performed outside the collision set. We deduced that  for $\alpha<2$ the two flows are topologically equivalent, and  showed that  this is not the case for $\alpha \geq 2.$ For the latter situation, we introduced the  idea of smoothing the amended potential and showed that, with such a modification, the two flows are topologically equivalent.

Smoothing of the amended potential in a two degree of freedom system might be considered a first approach to a more general problem: given a mechanical system with symmetry and with a singular potential,
what is the best way to apply smoothing? 
Our investigation shows that
the presence of non-inertial terms has to be treated carefully. Significant modifications of the global orbital picture might appear due to
 interplay of the potential and centrifugal  forces, especially when the system passes close to a degenerate (e.g. for $N$-body problems, a collinear) configuration. These issues will be discussed elsewhere.

\section*{Acknowledgements}
The authors thank to Andreea Font for helpful comments. This work was supported by the  NSERC,  Discovery Grants Program.



\begin{thebibliography}{2007}

\bibitem {aaseth-phd}S. J. Aarseth [1963], Dynamics of galaxies,  PhD Thesis, University of Cambridge. 


\bibitem {aaseth}  S. J. Aarseth [1963], Dynamical evolution of clusters of galaxies I,  {\it Mon. Not. R. Astron. Soc.}, $\bf 126$, 223-255. 

\bibitem{Arnold} V.I. Arnold  [1978],  {\it Mathematical methods in classical mechanics}, Springer-Verlag.




\bibitem{bellettini} G. Bellettini, G. Fusco, G.F. Gronchi [2003], Regularization of the two body problem via Smoothing of the Potential, {\it Communications on Pure and Applied Analysis}, $\bf{3}$, 317-347.
\bibitem{degiorgi} E. De Giorgi [1995], Congetture riguardanti alcuni problemi di evoluzione, {\it Duke Math. J.}, $\bf{81}$, 255-268.
\bibitem{dyer}
C.C. Dyer and P.S.S Ip [1993], Softening in $N-$body simulations of collisionless systems, {\it The Astronomical Journal}, $\bf{409},$  60-67.

\bibitem{golstein} H. Goldstein  [1980],  {\it Classical Mechanics}, Addison-Wesley, Series in Physics, Second Edition.

\bibitem{mcgehee}
R. McGehee  [1981], Double Collisions for a classical particle system with nongravitational interactions, {\it Comment. Math. Helvetici}, $\bf 56$, 524--557.

\bibitem {MP03}[MP03]  S. L. W. McMillan and S. F. Portegies Zwart [2003], The fate of star clusters near the galactic center. I. Analytic considerations   {\it The Astrophysical Journal}, $\bf 596$, 314-322. 

\bibitem{merritt}
D. Merritt [1996], Optimal Smoothing for N-body Codes, {\it The Astronomical Journal}, $\bf 111$, 2462-2464. 

\bibitem{neunzert}
H. Neunzert [1980], An introduction to the nonlinear Boltzmann-Vlasov equation, in Kinetic Theories and the Boltzmann Equation, Lecture Notes in Math., Springer-Verlag, $\bf 1048$, 60-110.


\bibitem{stoica}
C. Stoica [2002], Classical Scattering and Block
Regularization for the Homogeneous Central Field Problem,
{\it Celestial Mechanics and Dynamical Astronomy}, {\bf 84}, No.3,
223-229.


\end{thebibliography}
\end{document}